\documentstyle[seceq,supplement,epsf,wrapft,amsbsy]{ptptex}

\newcommand{\bra}[1]{\langle {#1} |}
\newcommand{\ket}[1]{| {#1} \rangle}
\newcommand{\vecr}{{\bf r}}

\notypesetlogo  

\title{
3D Real-Space Calculation of the Continuum Response
}

\author{
Takashi {\sc Nakatsukasa} 
 and Kazuhiro {\sc Yabana}$^{*,}$
}

\inst{
Physics Department, Tohoku University, Sendai 980-8578, Japan
\\
$^*$Institute of Physics, 
University of Tsukuba, Tsukuba 305-8571, Japan}

\recdate{
}

\abst{
We present that a linear response theory in the continuum
can be easily formulated with Absorbing Boundary Condition (ABC).
The theory is capable of describing continuum spectra
and dynamical correlations.
Application of the ABC does not require the spherical symmetry and
the method is suitable for mesh representation in the real coordinate space.
Isovector giant dipole resonances in beryllium isotopes are studied with
the time-dependent Hartree-Fock with the Skyrme force
in a three-dimensional mesh space with the ABC.
}

\begin{document}

\maketitle

\section{Introduction}
The drip-line nuclei are finite fermion systems whose separation energy
is nearly zero.
For such weakly bound systems, the continuum should be properly
taken into account in description of their structures and reactions.
In the linear response regime,
the inclusion of the single-particle continuum for particle-hole (p-h)
excitations has been achieved by using a method, so called, Continuum
Random-Phase Approximation (CRPA).\cite{SB75}
The trick to treat the continuum was use of the single-particle Green's
function with the Outgoing Boundary Condition (OBC).
This relies on explicit calculations of Green's functions with the OBC
in the one-dimensional (radial) coordinate.
We have developed a method to calculate the continuum response
in the three-dimensional (3D) real space by solving equations iteratively
with the Green's functions.\cite{NY01_1,NY01_2}
The OBC is explicitly taken into account with this method.
However, it is costly in computer resources.
We have also been investigating a possibility to approximate the OBC by
Absorbing Boundary Condition (ABC).\cite{NY01_1,NY01_2,NY02}
An advantage of the ABC is its simplicity and flexibility.
For instance, it is extremely difficult to treat the OBC in
the real-time calculation of the time-dependent wave functions,
while it is straightforward to implement the real-time method
with the ABC.\cite{NY01_1}

The Skyrme Hartree-Fock (HF) theory \cite{VB72}
has been extensively applied to
study of ground-state properties of relatively heavy nuclei.
To investigate their excited states,
a straightforward extension is
the linear response calculation based on the HF ground state.
The CRPA combined with the Skyrme Hartree-Fock (HF) theory
has been extensively utilized to study giant resonances
in spherical nuclei.\cite{GS81,HSZ96,HS96}
So far, only spherical nuclei were studied because of the limitation
of the conventional CRPA.
We would like to show that the linear response calculation with ABC
is a feasible extension of the {\it spherical} CRPA to the
{\it deformed} one.

\section{Skyrme Time-Dependent Hartree-Fock (TDHF) with ABC}
\label{sec: method}

Since the Hamiltonian in the Skyrme HF theory is almost diagonal in
coordinate representation, a mesh representation in the 3D coordinate space
provides an economical description.
The real-time calculation of the Skyrme TDHF were carried out
for studies of heavy-ion reactions.\cite{Neg82}
The main issue is then how to incorporate the proper boundary condition
in the 3D uniform grid representation.
In this paper,
we adopt a method of ABC combined with the real-time TDHF.\cite{NY01_1}

As is well known,
the Green's function with OBC (for $E>0$) can be simply written as
\begin{equation}
G^{(\sc obc)}(\vecr,\vecr'; E) = \bra{\vecr}\frac{1}{E-H+i\eta}\ket{\vecr'} ,
\end{equation}
where $\eta$ is a positive infinitesimal.
In numerical calculations, we take a spherical box with
radius $R$ as a model space.
Since the infinitesimal $\eta$ cannot be treated directly
in actual numerical calculation, we need to explicitly construct the outgoing
solutions at the boundary ($r\ge R$).
This is a difficult task when the system does not possess
a spherical symmetry.
The idea of ABC is to replace the $\eta$ by a finite coordinate-dependent
quantity,
$\bra{\vecr}\tilde\eta\ket{\vecr'}=\tilde\eta(\vecr)\delta(\vecr-\vecr')$,
and imposes the vanishing boundary condition:
\begin{equation}
G^{(\sc abc)}(\vecr,\vecr'; E)
 = \bra{\vecr}\frac{1}{E-H+i\tilde\eta}\ket{\vecr'} .
\end{equation}
Here, $\tilde\eta(\vecr)$ is taken to be positive far outside the system
($R<r<R+\Delta r$) and zero elsewhere.
The Green's function has a vanishing boundary condition,
$G^{(\sc abc)}(\vecr,\vecr'; E)=0$
for $r=R+\Delta r$ or $r'=R+\Delta r$.
This is equivalent to bound-state Green's function
for the Hamiltonian with
a complex absorbing potential, $H-i\tilde\eta(\vecr)$.
Now let us discuss properties of $G^{(\sc abc)}(E)$.

For $E<0$, the Green's function with ABC has a pole at
eigenenergies $E=E_n$, because the bound eigenstate is localized,
$\psi_n(\vecr)\approx 0$ at $r>R$.
For $E>0$, the Green's function has an outgoing asymptotic behavior.
This can be seen as follows:
In the outer region of our model space ($r>R$),
the complex potential $-i\tilde\eta$ produces
complex wave numbers,
$k+i\gamma$ where $\gamma>0$.
Thus, the outgoing waves exponentially damp while incoming waves diverge.
The vanishing boundary condition selects only the outgoing solutions.
In the inner model space ($r<R$),
as far as the reflection caused by the potential,
$-i\tilde\eta(\vecr)$, is negligible,
$G^{(\sc abc)}(E)$ is identical to $G^{(\sc obc)}(E)$.
Therefore, 
the complex boundary potential must be strong enough
to absorb the whole outgoing wave and simultaneously gentle enough
to minimize the reflection.
We adopt an absorbing potential of linear dependence on
the radial coordinate,
\begin{equation}
\label{eta_tilde}
\tilde\eta(r) =
\cases{
0 ,                         &  \mbox{for $r < R$,} \cr
\eta_0 \frac{r-R}{\Delta r},   &  \mbox{for $R < r < R+\Delta r$.}\cr
}
\end{equation}
In order to minimize the reflection,
the height $\eta_0$ and width $\Delta r$ should satisfy
a condition\cite{Chi91,NY01_1}
\begin{equation}
\label{good_absorber}
20\frac{E^{1/2}}{\Delta r \sqrt{8m}} < \eta_0
< \frac{1}{10} \Delta r \sqrt{8m} E^{3/2} .
\end{equation}

An advantage of the ABC is its applicability to the real-time
TDHF calculations.
We use the Skyrme TDHF method to investigate giant resonances in the
continuum.
The initial state at $t=0$ is prepared by boosting
the HF ground state, $\ket{\Psi_0}$, with an instantaneous external field,
$V_{\rm ext}(\vecr,t)=v_{\rm ext}(\vecr)\delta(t)$.
Then, the time evolution of the TDHF sate, $\ket{\Psi(t)}$,
in a 3D real space is calculated
according to a prescription given in Ref.~\citen{FKW78}.
Time step is taken as $\Delta t=0.001\mbox{ MeV}^{-1}$ and
$N$-times iteration of time evolution is done for $\ket{\Psi(t)}$
up to the total period $T=N\Delta t$.
After the time evolution,
we carry out the Fourier transform of
expectation values of a one-body operator $\hat{D}$,
$\bra{\Psi(t)}\hat{D}\ket{\Psi(t)}$.
The imaginary part of this quantity gives the corresponding excitation
strength,
$S(\hat{D};E)=\sum_n |\bra{n}\hat{D}\ket{0}|^2\delta(E-E_n)$.
See Ref.~\citen{NY01_1} for details of the numerical calculation.

\section{Numerical results and discussions}

        \begin{wrapfigure}{r}{6.6cm}
            \epsfxsize = 6.6cm   
            \centerline{\epsfbox{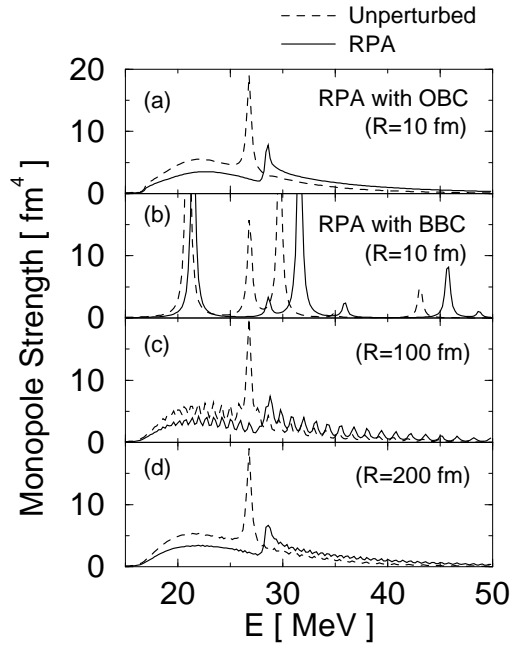}}
        \caption{Strength functions of giant monopole resonances in $^{16}$O
                 calculated with the BKN interaction.
                 A smoothing parameter, $\Gamma=0.5$ MeV, is added to the
                 energy as $E+i\Gamma$.
                 (a) Results of CRPA using a radial coordinate
                     up to $R=10$ fm.
                 (b) RPA with the BBC of $R=10$ fm.
                 (c) The same as (b) except for $R=100$ fm.
                 (d) The same as (b) except for $R=200$ fm. }
        \label{fig: Cont-Discr}
        \end{wrapfigure}
\subsection{Continuum effects: an illustration}

In many cases, the RPA response functions have been calculated
utilizing the $L^2$ basis set, such as harmonic oscillator basis.
A similar $L^2$-type calculation is possible using the coordinate space
of finite box with the vanishing boundary condition.
We call this ``Box Boundary Condition (BBC)''.
The low-energy RPA response below the separation energy
are well described with the BBC.
Then, in order to obtain a reasonable BBC description of
the response in a energy region above the separation energy,
what size of model space is required?
In other words, how large the model space should be, to simulate
the continuum wave functions?

Assuming the BBC of box radius $R$,
an escaping particle of velocity $v$ (energy $E$)
reflects at the boundary wall at $r=R$.
This reflection gives a spurious contribution to
the continuum response of the system and
eventually produces a discrete spectrum.
Thus, the maximum period of {\it physical} time evolution is
\begin{equation}
T=R/v=R\sqrt{M/2E} .
\end{equation}
The uncertainty principle tells that
the total period $T$ is related to
an energy resolution $\Delta E$ after the Fourier transform:
\begin{equation}
\Delta E = \frac{2\pi}{T} = \frac{2\pi}{R}\sqrt{\frac{2E}{M}} .
\end{equation}
Here, $M$ and $E$ are the mass and the kinetic energy of
escaping particles, respectively.
Therefore, for instance, if we require $\Delta E=1$ MeV for
the strength function at $E=10$ MeV above the threshold,
one has to handle the box of $R\approx 200$ fm.
Dealing with this large space is difficult even for a spherical (1D) case
and is almost impossible for a deformed (3D) case.

Using a simplified Skyrme force, the BKN interaction,\cite{BKN76}
monopole response in $^{16}$O is calculated
both with and without the continuum.
For this case,
owing to the simplicity of the BKN interaction and
the spherical symmetry of the HF state,
we may carry out the RPA calculation with the BBC
in a large coordinate space.
We also perform the CRPA calculation.
The OBC calculation is done for a spherical box of $R=10$ fm,
the result of which is compared with the BBC results with
three different boxes of $R=10$, 100, and 200 fm
(Fig.~\ref{fig: Cont-Discr}).
The figure shows that we need utilize a coordinate space of
$R>100$ fm to obtain a sensible result.
On the other hand, if we use the complex potential of BBC,
Eq.~(\ref{eta_tilde}), a space of $\Delta r=10$ fm in addition to
$R=10$ fm is enough to obtain an identical result to the OBC.\cite{NY02}

\subsection{Giant dipole resonances in beryllium isotopes}

Isovector dipole states in even beryllium (Be) isotopes
($^{8-14}$Be) are investigated using
the real-time-ABC method in Sect.\ref{sec: method}.
We adopt the SIII parameters for the Skyrme interaction including
all time-even densities and most of time-odd densities.\cite{BFH87}
Properties of the HF ground states of these isotopes
were investigated in Ref.~\citen{TYI95}.
The ground-state binding energies are well reproduced with the SIII
and prolate shapes of the density distributions are obtained for
$^{8,10,14}$Be.
We now investigate the $E1$ linear response on these ground states.
An initial state for the real-time TDHF calculation
is prepared by boosting the ground state with
an $E1$ operator with a recoil charge,
$v_{\rm ext} = (Z/A)\boldsymbol{r}_n - (N/A)\boldsymbol{r}_p$.
The total period of time evolution is taken as $T=4\mbox{ MeV}^{-1}$.
This corresponds to about an energy resolution of 1.6 MeV.
The model space we take is a 3D lattice coordinate space of $R=8$ fm
with a square mesh of $\Delta x =\Delta y=\Delta z=1$ fm.
For the ABC potential, Eq.~(\ref{eta_tilde}),
we take $\eta_0=14$ MeV and $\Delta r=10$ fm.

\begin{figure}
            \epsfxsize = 10cm   
            \centerline{\epsfbox{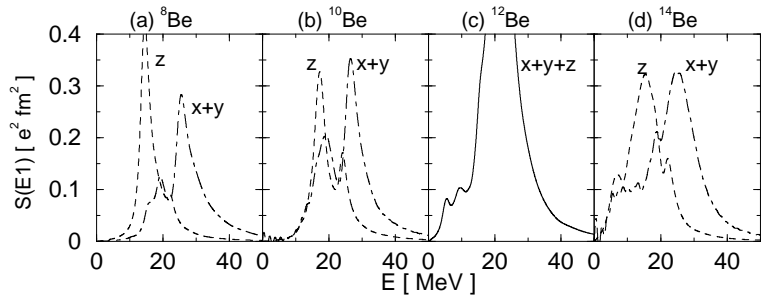}}
        \caption{$E1$ strength functions of giant dipole resonances in
                 even-$A$ Be isotopes, calculated by
                 the TDHF with the SIII force using the ABC potential:
                 $R=8$ fm, $\Delta r=10$ fm, $\eta_0=14$ MeV.
                 $E1$ strengths parallel (perpendicular) to the symmetry
                 axis are shown by dashed (dot-dashed) lines.
                 }
        \label{fig: Be_E1}
        \end{figure}

In Fig.~\ref{fig: Be_E1},
the obtained $E1$ strength functions are presented.
Since $^{8,10,14}$Be have prolate shapes in the ground states,
the $E1$ strengths parallel to the symmetry ($z$) axis and perpendicular
($x$ and $y$) axes are displayed separately.
Large deformation splitting over 10 MeV is obtained in these nuclei.
This is because the ground states have large deformations which almost
correspond to {\it superdeformation} in heavy nuclei.
The ground state of $^{12}$Be is nearly spherical and we show the
$E1$ response summed with respect to three directions
in Fig.~\ref{fig: Be_E1} (c).

Appearance of low-energy (pigmy) strengths ($E<10$ MeV) is seen
in $^{12}$Be and even enhanced in $^{14}$Be.
These strengths disappear if we get rid of the ABC potential,
which suggests that the continuum is essential to produce the low-energy
$E1$ strengths near threshold.
It is also worth noting that the real part of the response function
changes its sign at the position of the low-energy peak.
This may suggest a real resonance character.

\section{Conclusions}
In this paper, we present the method of ABC which is able to describe
the 3D continuum effects in a feasible manner and in good accuracy.
We take $E1$ giant resonances in beryllium isotopes and apply the
real-time TDHF-ABC method.
Large deformation splittings in $^{8,10,14}$Be and low-energy
resonances in $^{12,14}$Be are obtained.

Although we have shown only the application to the response calculations
in this paper,
the ABC provides a general and powerful method to treat the OBC.
Another application,
a calculation of breakup processes of a halo nucleus,
is also under progress.\cite{Yab02}
We are encouraged by the results we have obtained and will
study structures and reactions of unstable nuclei using the ABC.

\end{document}